\documentclass[twocolumn]{aastex62}

\newcommand{\pisco}{J1342+0928}

\newcommand{\kms}{{\rm km\,s}\ensuremath{^{-1}}}

\newcommand{\cii}{[\ion{C}{2}]}

\newcommand{\mgii}{\ion{Mg}{2}}

\graphicspath{{./}{figures/}}

\accepted{ApJ Letters, July 29, 2019}

\shorttitle{\cii\ high-resolution dynamics from a $z=7.54$ quasar}
\shortauthors{Ba\~nados et al.}

\begin{document}


\title{\bf The  $z=7.54$ Quasar ULAS~\pisco\ Is Hosted by a Galaxy Merger}

\correspondingauthor{Eduardo Ba\~nados}
\email{banados@mpia.de}

\author[0000-0002-2931-7824]{Eduardo Ba\~nados}
\affiliation{Max-Planck-Institut f\"{u}r Astronomie, K\"{o}nigstuhl 17, D-69117, Heidelberg, Germany}
\affiliation{The Observatories of the Carnegie Institution for Science, 813 Santa Barbara St., Pasadena, CA 91101, USA}

\author[0000-0001-8695-825X]{Mladen Novak}
\affiliation{Max-Planck-Institut f\"{u}r Astronomie, K\"{o}nigstuhl 17, D-69117, Heidelberg, Germany}

\author[0000-0002-9838-8191]{Marcel Neeleman}
\affiliation{Max-Planck-Institut f\"{u}r Astronomie, K\"{o}nigstuhl 17, D-69117, Heidelberg, Germany}

\author[0000-0003-4793-7880]{Fabian Walter}
\affiliation{Max-Planck-Institut f\"{u}r Astronomie, K\"{o}nigstuhl 17, D-69117, Heidelberg, Germany}
\affiliation{National Radio Astronomy Observatory, Pete V. Domenici Array Science Center, P.O. Box 0, Socorro, NM 87801, USA}

\author[0000-0002-2662-8803]{Roberto Decarli}
\affiliation{INAF -- Osservatorio di Astrofisica e Scienza dello Spazio, via Gobetti 93/3, I-40129, Bologna, Italy}

\author[0000-0001-9024-8322]{Bram P. Venemans}
\affiliation{Max-Planck-Institut f\"{u}r Astronomie, K\"{o}nigstuhl 17, D-69117, Heidelberg, Germany}

\author[0000-0002-5941-5214]{Chiara Mazzucchelli}
\affiliation{European Southern Observatory, Alonso de Cordova 3107, Vitacura, Region Metropolitana, Chile}

\author[0000-0001-6647-3861]{Chris Carilli}
\affiliation{National Radio Astronomy Observatory, Pete V. Domenici Array Science Center, P.O. Box 0, Socorro, NM 87801, USA}
\affiliation{Astrophysics Group, Cavendish Laboratory, JJ Thomson Avenue, Cambridge CB3 0HE, UK}

\author[0000-0002-7633-431X]{Feige Wang}
\affiliation{Department of Physics, Broida Hall, University of California, Santa Barbara, CA 93106--9530, USA}

\author[0000-0003-3310-0131]{Xiaohui Fan}
\affiliation{Steward Observatory, The University of Arizona, 933 North Cherry Avenue, Tucson, AZ 85721--0065, USA}

\author[0000-0002-6822-2254]{Emanuele~P.~Farina}
\affiliation{Department of Physics, Broida Hall, University of California, Santa Barbara, CA 93106--9530, USA}

\author[0000-0003-4996-9069]{Hans-Walter Rix}
\affiliation{Max-Planck-Institut f\"{u}r Astronomie, K\"{o}nigstuhl 17, D-69117, Heidelberg, Germany}

\begin{abstract}
We present ALMA $0\farcs28 \times 0\farcs20$ ($1.4\,\mathrm{kpc} \times 1.0\,\mathrm{kpc}$) resolution observations of the \cii\ $158\,\mu$m line and underlying dust continuum in the host galaxy of the most distant quasar currently known, ULAS~\pisco\ at $z=7.5413$. Both the \cii\ and continuum are detected and spatially resolved. The \cii\ line is $\sim$1.5 times more extended than the continuum emission, showing an elongated and complex
structure of approximately $3.2\,\mathrm{kpc}\times 6.4\,\mathrm{kpc}$.
Two separate peaks are clearly seen ($\gtrsim$$6\sigma$ each) in three 100\,\kms\ width \cii\ channel maps.
The  velocity field of the \cii\ gas does not show evidence of a coherent rotation field but rather chaotic motion reminiscent of an ongoing merger.

\end{abstract}

\keywords{cosmology: observations --- cosmology: early universe
--- quasars: individual (ULAS J134208.10+092838.35)
--- galaxies: ISM
--- galaxies: kinematics and dynamics}


\section{Introduction} \label{sec:intro}

The host galaxies of high-redshift quasars are thought to be the most massive galaxies at the earliest cosmic epochs, and therefore provide important constraints on early galaxy formation and evolution models. The physical properties of these extreme galaxies can be studied in detail with existing (sub)millimeter facilities even within the first billion years of the universe ($z>6$;  see \citealt{carilli2013} for a review).
Pioneering works\ pushed the limits of facilities such as PdBI\footnote{Plateau de Bure Interferometer.}/NOEMA\footnote{NOrthern Extended Millimeter Array.} and the VLA\footnote{Karl G. Jansky Very Large Array.} to characterize the dust and the molecular and atomic gas in a handful of the brightest of these quasars (e.g., \citealt{bertoldi2003, walter2004, maiolino2005, walter2009, venemans2012,banados2015b}). These studies showed that early supermassive black hole growth can be accompanied by extended, intense star formation and large reservoirs of dense and enriched molecular gas.

The \cii\ $158\,\mu$m fine-structure line (hereafter \cii) is one of the brightest far-infrared lines in star-forming galaxies, and its frequency at $z>6$ is conveniently located in a high-transmission atmospheric window visible with millimeter facilities.  To date, thanks to the advent of ALMA\footnote{Atacama Large Millimeter/submillimeter Array.} more than 40 quasar host galaxies at $z\gtrsim 6$ have been detected in \cii\ \cite[e.g.,][]{willott2017, izumi2018, decarli2018} and a handful of these objects have (sub)kiloparsec-resolution maps of their interstellar medium (e.g., \citealt{venemans2016, venemans2019, wang-ran2019, neeleman2019}).
These high-resolution observations are exposing a diverse population. For example, recent $\sim$$1\,$kpc resolution ALMA \cii\ imaging of the quasar ULAS~J1120+0641 at $z=7.1$ revealed an extremely compact, unresolved ($\lesssim 1\,$kpc) host galaxy that does not exhibit ordered motion on kiloparsec scales \citep{venemans2017a}, in contrast to what is observed in other $z\sim 6$ quasar hosts \citep[e.g.,][]{shao2017}.

In this Letter we present kiloparsec-resolution observations of the \cii\ line and underlying dust continuum of the host galaxy of the quasar ULAS~\pisco\ at $z=7.54$ (hereafter \pisco). In a companion paper we present a multiline ALMA survey of this quasar host galaxy, tracing various phases of its interstellar medium \citep{novak2019}.
This is the most distant quasar known to date and is powered by a supermassive black hole with 800 million times the mass of the sun \citep{banados2018a}.   This system is being studied from X-rays to radio wavelengths (e.g., \citealt{banados2018b}; E.~Momjian et al.\ 2019, in preparation). IRAM/NOEMA observations resulted in the detection of the \cii\ line and dust from this quasar at $z=7.5413\pm0.0007$, constraining the far-infrared luminosity of the quasar to $(0.5-1.4)\times 10^{12}\,L_\odot$. The NOEMA \cii\ emission remained spatially unresolved with an upper limit on the diameter of 7\,kpc \citep{venemans2017c}.

We assume a standard, flat, $\Lambda$-CDM cosmology with $\Omega_\Lambda = 0.7$,  $\Omega_m = 0.3$,  and $H_0 = 70~\text{km}~\text{s}^{-1}~\text{Mpc}^{-1}$. In this cosmology, at $z=7.54$ the universe was 680 Myr old and $0\farcs2$ corresponds to a projected physical separation of 1\,kpc.

%




\section{Observations} \label{sec:observations}

We performed ALMA Band 6 observations to spatially resolve the \cii\ line of \pisco. The observations were carried out on 2017 December 26 in the C43-6 array with a total on-source time of 114 minutes. The observations were tuned to cover 222.511\,GHz, which corresponds to the frequency of the \cii\ line ($\nu_{\rm rest}= 1900.5369\,$GHz) at the redshift of the quasar, $z=7.5413$, measured from  NOEMA data \citep{venemans2017c}. According to the ALMA documentation, the maximum recoverable scale\footnote{Defined as the largest angular size at which at least 10\% of the total flux density of a uniform disk is recovered, and estimated using the 5th percentile of the sampled $uv$ distances.} of our observations is 3$\arcsec$.

We used the default ALMA calibration pipeline implemented in
the Common Astronomy Software Application package \citep[CASA; ][]{mcmullin2007} to calibrate the dataset\footnote{CASA calibration pipeline version 5.1.1-5}. Visibilities were imaged with the {\sc tclean}\footnote{CASA version 5.4.0-68} task using natural weighting to maximize the signal-to-noise (S/N) of our detection, and cleaning was performed down to $2\sigma$ inside a circle with a radius of $2\arcsec$ centered at the source. The continuum was subtracted from visibilities with the task {\sc uvcontsub} to facilitate imaging of the continuum-free data, such as the \cii\ line channel and moment zero maps, as well as to obtain the pure continuum data.
The imaged cube has a channel width of 50\,\kms, synthesized beam of $0\farcs28 \times 0\farcs20$, and an rms noise level of 0.1\,mJy\,beam$^{-1}$ for each channel. An additional cube with a broader channel width of 100\,\kms\ (rms of 0.07\,mJy\,beam$^{-1}$) was imaged to assist in the analysis of the source morphology (Section \ref{sec:chmaps}).
Data containing only the continuum were imaged yielding an rms value of 7.7\,$\mu$Jy\,beam$^{-1}$ in the final map.

\begin{figure}[t!]
\plotone{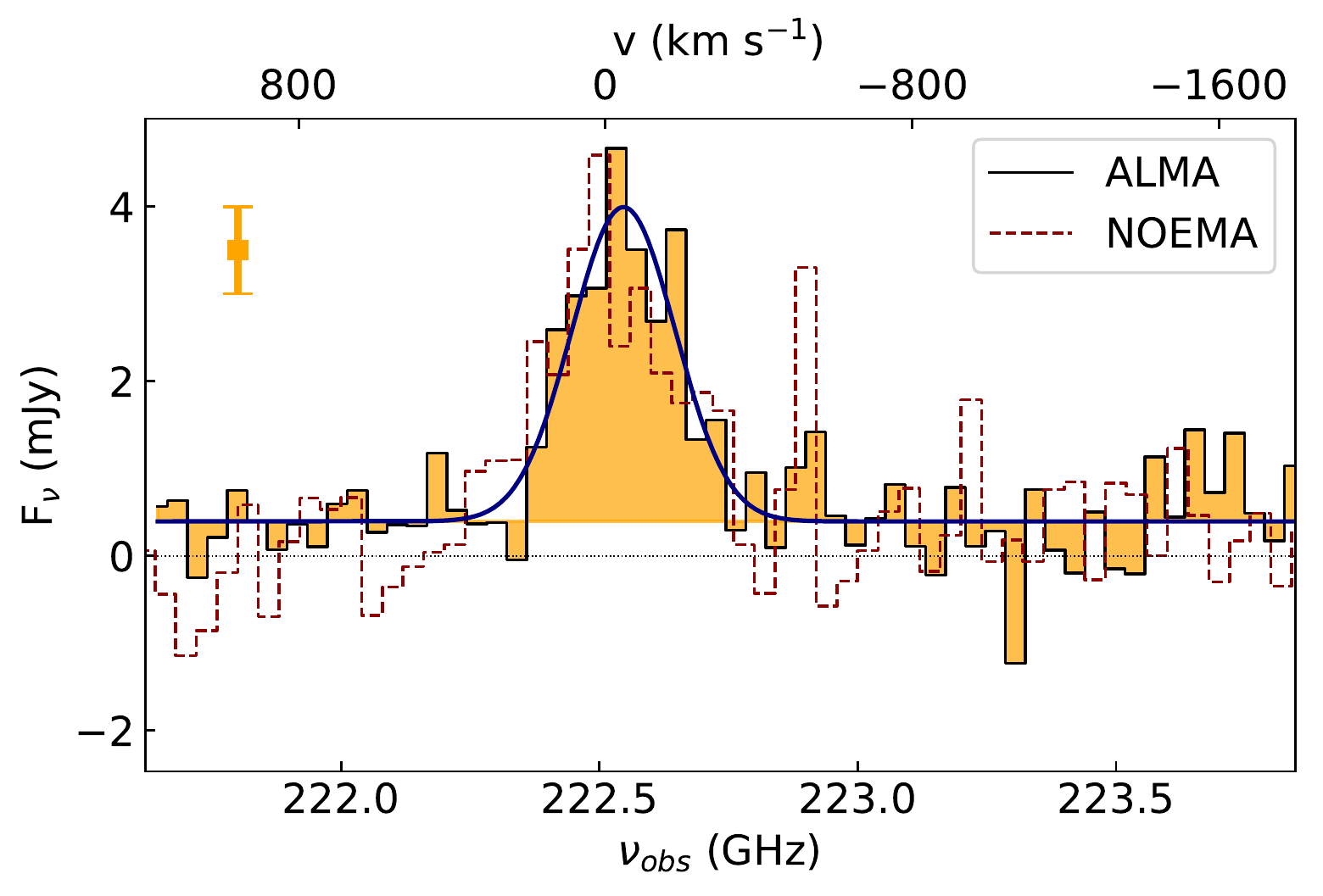}
\caption{Spatially integrated \cii\ spectrum of the quasar \pisco\ within an aperture of $1\farcs3$ radius.
Standard deviation of the flux measurement is shown on the left.
The blue line is a fit to the data consisting of a Gaussian on top of a flat continuum.  The upper axis shows velocities centered at $z=7.5413$ as reported in \cite{venemans2017c},  which is consistent with the value measured here from the ALMA data: $z=7.5400\pm 0.0003$.  Their NOEMA spectrum is also shown with a dashed line.
\label{fig:spectrum}}
\end{figure}







\begin{figure*}[ht!]
\plottwo{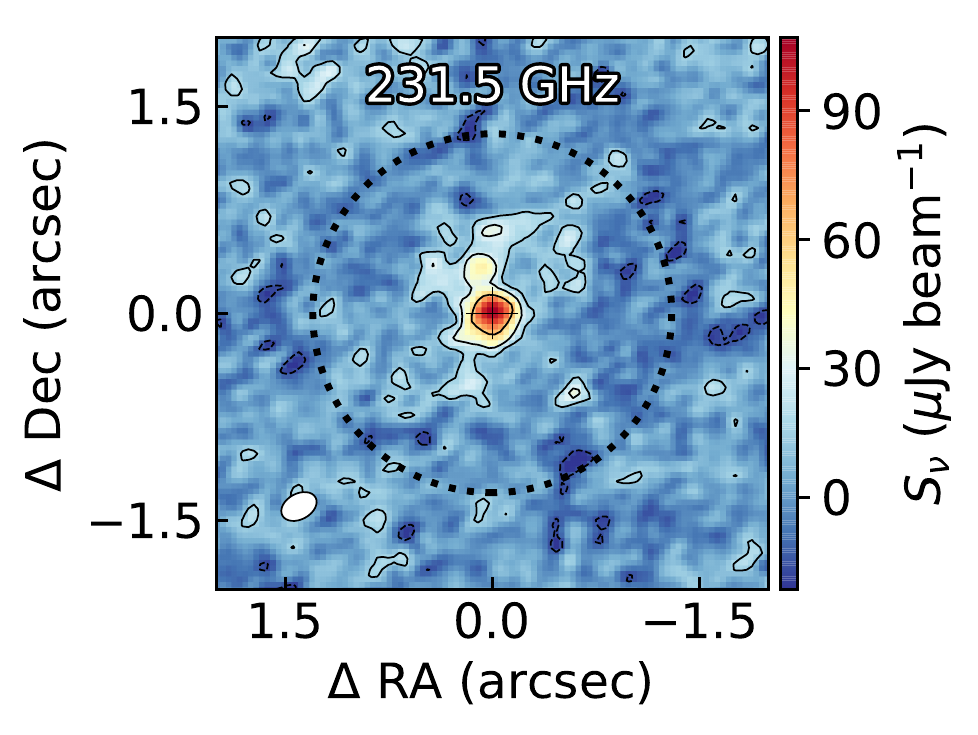}{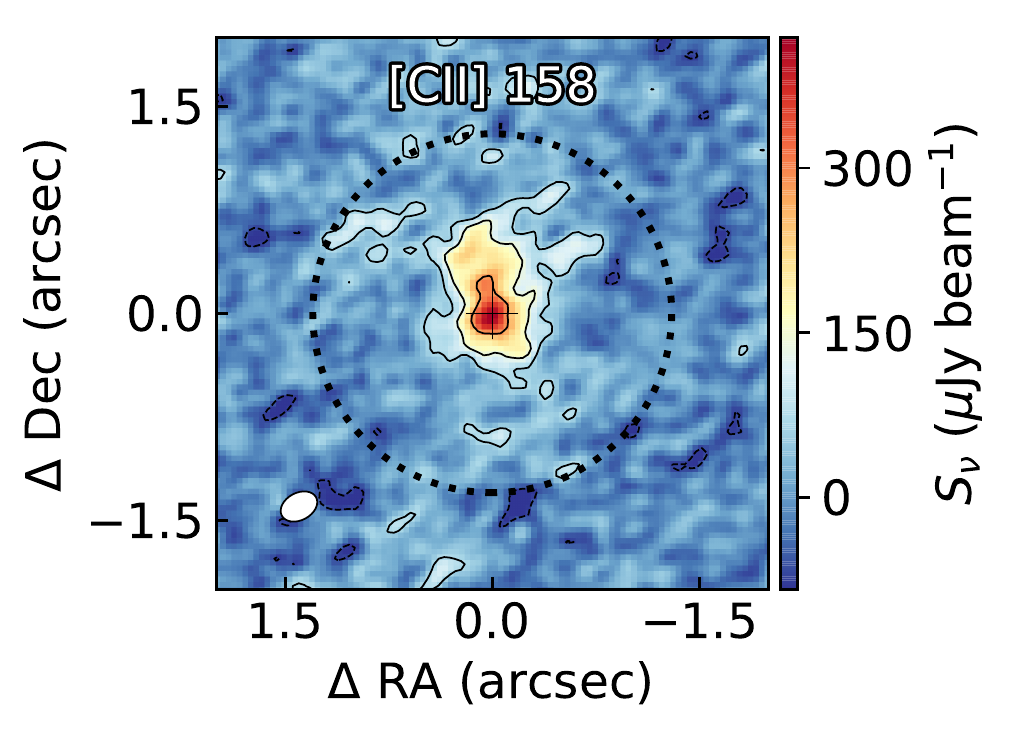}
\caption{ALMA dust continuum (left) and \cii\ (right) maps of \pisco. The $0\farcs28 \times 0\farcs20$ ($1.41\,\mathrm{kpc} \times 0.99\,\mathrm{kpc}$) beam size is shown at the bottom left of each panel while the cross indicates the peak of the dust continuum emission, which is consistent with the near-infrared position of the quasar. The solid (dashed) lines represent the $+$($-$)2, 4, 8 $\times \sigma$ contours with $\sigma=7.7\,\mu$Jy\,beam$^{-1}$ for the continuum flux density at  231\,GHz (left panel) and  $\sigma=35.0\,\mu$Jy\,beam$^{-1}$ for the \cii\ flux density (right panel).  The black dotted circles mark the $1\farcs3$ (6.5\,kpc) radius aperture used to measure the total flux densities.
\label{fig:alma-cont-cii}}
\end{figure*}

\section{Results and discussion} \label{sec:discussion}

\subsection{\cii\ spectrum}

To maximize the recoverable \cii\ emission spread over multiple beams, we extracted the spectrum using an aperture with a radius of $1\farcs3$ (6.5\,kpc), no further \cii\ is recovered beyond this radius (see Figures in Appendix A of \citealt{novak2019}).
In order to properly measure flux density of a resolved source we applied the residual scaling method, which mitigates the issue of ill-defined units in interferometric maps (see Appendix A in \citealt{novak2019}; and also \citealt{jorsater1995, walter1999} for further details).
The correct flux measurement is equal to $\epsilon\times D$, where $\epsilon=C / (D - R)$ is the clean-to-dirty beam area ratio, and $D$, $C$, and $R$ are aperture fluxes measured inside the dirty map, clean components only map, and the residual map, respectively. The scaling factor $\epsilon$ depends on the dirty beam shape and the size of the aperture used, and in our case is $\epsilon=0.4$.
In other words, we measure the flux in the dirty map and scale it by $\epsilon$, which ensures that the final flux density units are properly defined.
If we measured the total aperture flux in the final map, which is a combination of the clean Gaussian components on top of the residual, we would have obtained 1.25--1.8 larger values. The exact difference changes from channel to channel due to varying contributions from the clean and the dirty flux present inside the aperture.
The extracted spectrum is shown in Figure \ref{fig:spectrum}
and a Gaussian fit results in the integrated \cii\ line flux of $1.22\pm 0.15$\,Jy\,\kms, FWHM of $318 \pm 29$\,\kms, and a redshift of $7.5400 \pm 0.0003$. Recently, there has been growing evidence for significant velocity differences between the emission lines of the highest-redshift quasars \citep[e.g.,][]{mazzucchelli2017b,meyer2019}. Our ALMA data imply that the \mgii\ emission arising from the broad-line region of the quasar is blueshifted by $456\pm 140\,\kms$ with respect to the systemic redshift of the host galaxy traced by \cii.
These  measurements are consistent with the values reported in the lower-resolution NOEMA data \citep{venemans2017c}, and we use their published redshift value throughout the paper for consistency.
We note that the limitation of the maximum recoverable scale applies to smooth large-scale structures only. Patchy flux distribution within the aperture can still be accumulated. The agreement between our ALMA spectrum and the low-resolution ($\sim$$2\arcsec$) NOEMA spectrum (see Fig.~\ref{fig:spectrum}) demonstrates that we do not miss significant flux with our aperture choice.

\subsection{Continuum and \cii\ maps}

We imaged the line-free channels (effective bandwidth of 6.6\,GHz) to create the dust continuum map at 231.5\,GHz shown in the left panel of Figure \ref{fig:alma-cont-cii}, from which we measure a flux density of $0.39\pm 0.07$\,mJy in a circular aperture of radius $1\farcs3$, i.e., the same aperture used for the \cii\ spectrum shown in Figure \ref{fig:spectrum}.
We also imaged a moment zero \cii\ map over a 455\,\kms\ wide range, shown in the right panel of Figure \ref{fig:alma-cont-cii}.
This width corresponds to the 1.2$\times$FWHM measured in the low-resolution NOEMA data \citep{venemans2017c}, which was chosen to maximize the S/N of a Gaussian line detection\footnote{Choosing the FWHM of 318\,\kms, which we measure in our ALMA spectrum, yields  consistent results within the uncertainties.}.

Although the integrated spectrum does not show evident deviations from a simple Gaussian profile (Figure \ref{fig:spectrum}), the morphology and kinematics of this system are complex.
The continuum and \cii\ emission are both spatially resolved with approximate sizes of $2.3\,\mathrm{kpc}\times 4.0\,\mathrm{kpc}$ and $3.2\,\mathrm{kpc}\times 6.4\,\mathrm{kpc}$, respectively (Figure \ref{fig:alma-cont-cii}). The peak of the continuum emission at $13^{\rm h} 42^{\rm m} 08\fs098$ $+09^{\circ} 28^{\prime} 38 \farcs 35$ (ICRS) is consistent with the peak of the \cii\ emission as well as the near-infrared $J$-band position, within the uncertainty of 50 mas.

\subsection{Kinematics: Channel maps}
\label{sec:chmaps}
\begin{figure*}[ht!]
\plotone{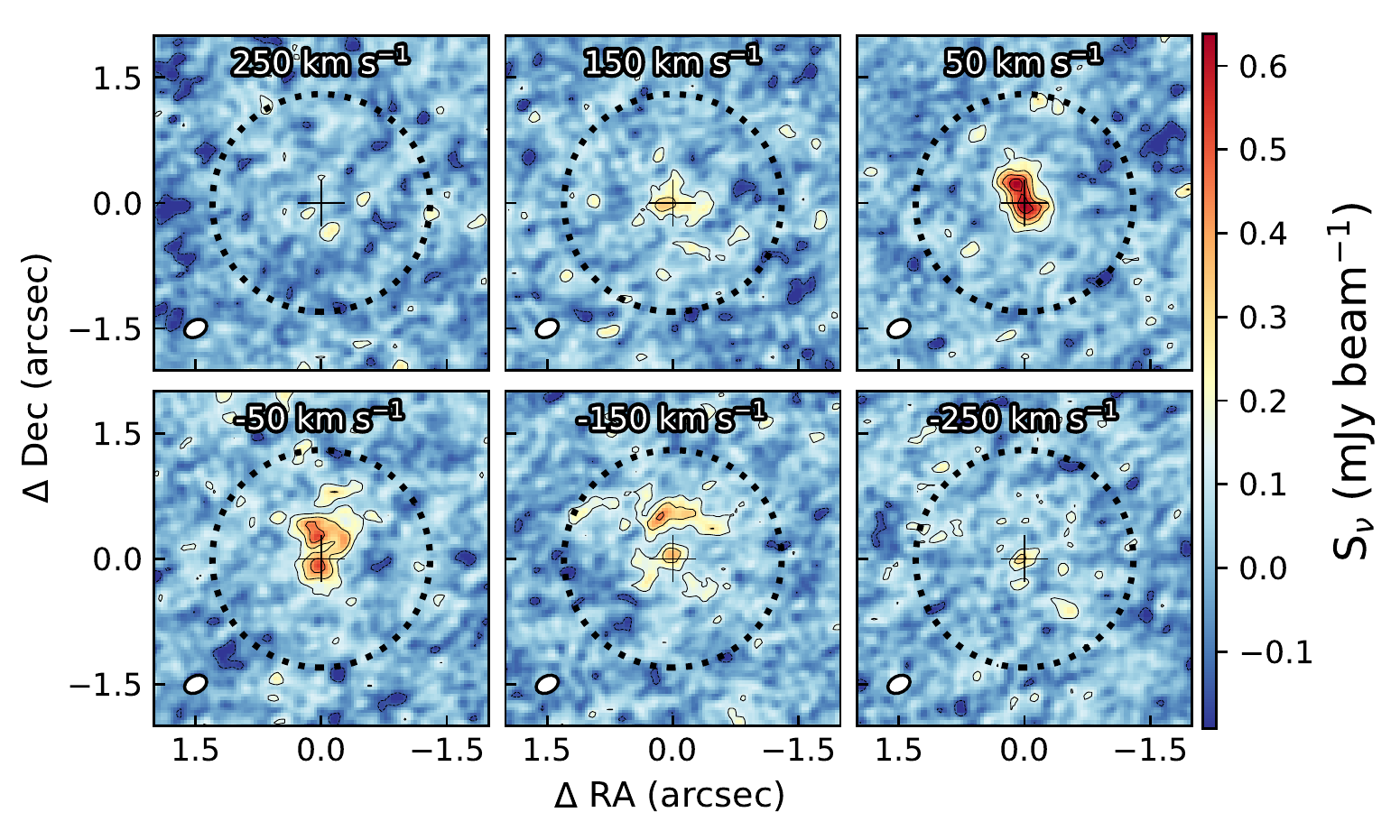}
\caption{\cii\ channel maps of the continuum-subtracted data cube of \pisco. The zero velocity is set to a redshift of $z=7.5413$. Solid (dashed) contours outline positive (negative) values starting at $2\sigma$ with steps of $2\sigma$, where $\sigma$ is the noise in each individual channel: $\sigma\approx0.07$\,mJy\,beam$^{-1}$. The synthesized beam is shown in the lower left corner. The black dotted circles are the apertures (r=$1\farcs3$) used to measure the total flux density (see Figure \ref{fig:alma-cont-cii}).
\label{fig:chmap}}
\end{figure*}

The channel maps of the continuum-subtracted data cube of \pisco\ are shown in Figure \ref{fig:chmap}.
\cii\ emission is detected over the entire width of the \cii\ line ($\sim$$320\,\kms$) at the position of the peak of the continuum emission (cross in Figure \ref{fig:chmap}).
This figure reveals a highly complicated \cii\ emission, being bright and complex in the central channel maps.
The emission breaks up into two structures with clear separate peaks, each detected at $\gtrsim 6\sigma$ significance in three different channels. We extracted the spectra centered on both structures using an aperture of 0\farcs25 (1.25\,kpc), resulting in spectra consistent with the total integrated \cii\ spectrum shown in Figure \ref{fig:spectrum}.
This resembles the clumpy substructures identified in some high-redshift submillimeter galaxies \cite[e.g.,][]{hodge2019} and quasar--galaxy mergers \cite[e.g.,][]{decarli2019}.
Higher S/N observations are required to reveal more details in narrower \cii\ channel maps and  higher resolution is needed to probe how clumpy these substructures are.

\subsection{\cii\ moment maps}
\begin{figure*}[ht!]
\plottwo{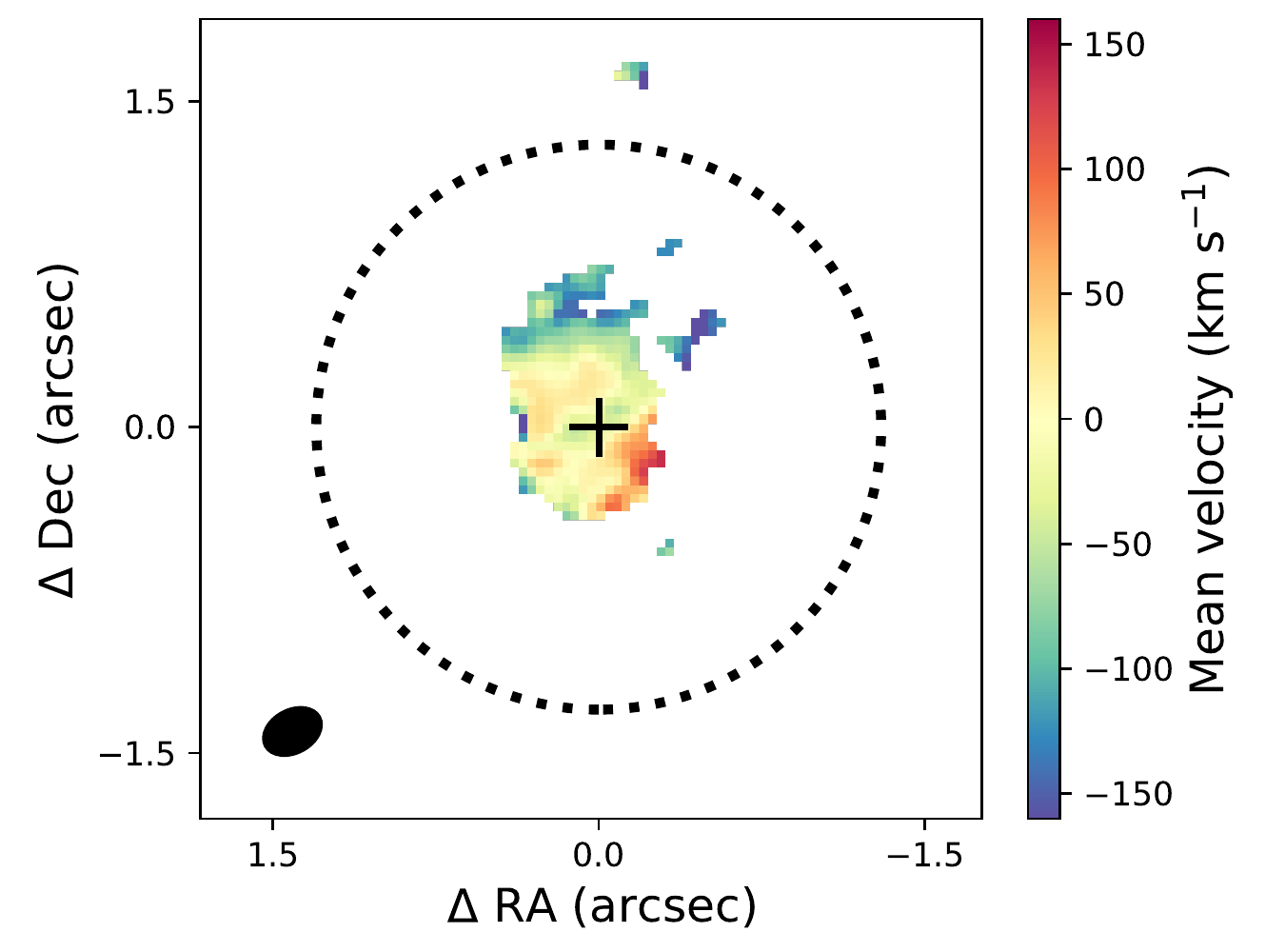}{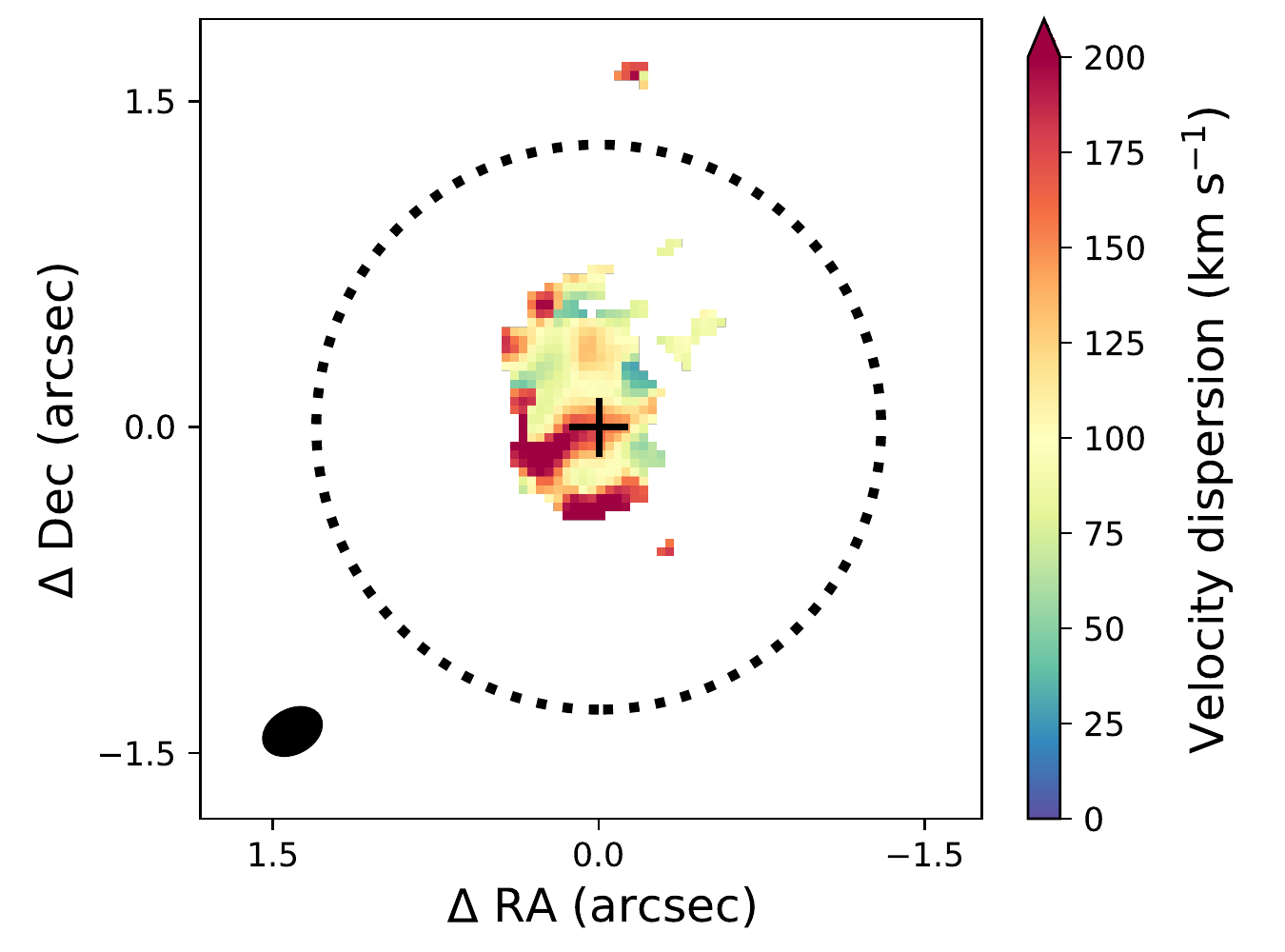}
\caption{
Mean velocity (left panel) and velocity dispersion (right panel) fields of the \cii\ line for \pisco. These quantities are estimated by fitting a Gaussian profile to the \cii\ spectrum of each individual pixel. Here the mean velocity is the location of the peak of the Gaussian fit, and the velocity dispersion is the square root of the variance of the Gaussian profile. Only those pixels that have a 3$\sigma$ detection in the integrated \cii\ flux (see Figure \ref{fig:alma-cont-cii}) are displayed.  The black dotted circles mark the region used to measure the total flux densities (see Figure \ref{fig:alma-cont-cii}), and the synthesized beam is shown in the bottom left. The black cross marks the position of the quasar.
\label{fig:moments}}
\end{figure*}

The velocity field and velocity dispersion of the \cii\ line are shown in Figure \ref{fig:moments}.  Even though there seems to be a mild velocity gradient with the northern part blueshifted and the southern part redshifted, the velocity dispersion does not resemble a coherent rotating structure as seen in some quasars at $z\sim 6$ \citep[e.g.,][]{shao2017}. Indeed, the velocity dispersion has a peak at the quasar location in line with what observed in the channel maps (Figure \ref{fig:chmap}).

\subsection{Rotating disk model}
\label{sec:model}
\begin{figure*}[ht!]
\plotone{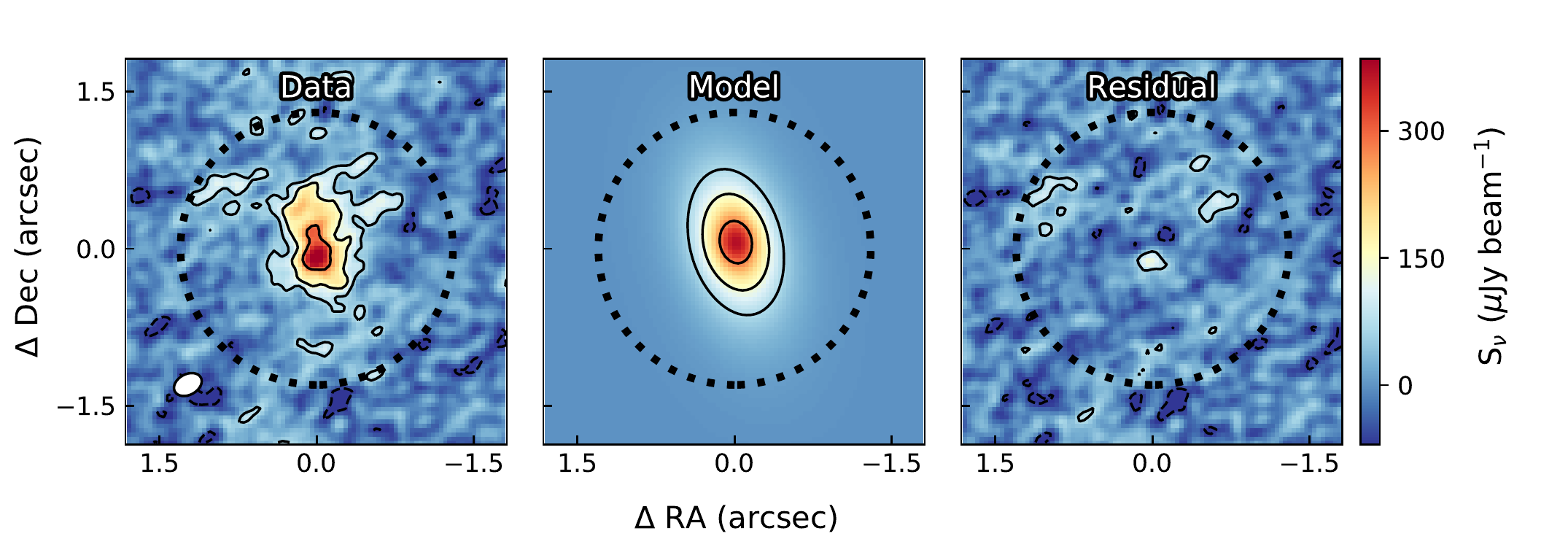}
\caption{The left panel is the moment zero of the data (as in the right panel of Figure \ref{fig:alma-cont-cii}). The middle panel is the best-fit model of a simple smooth disk to the data (see Section \ref{sec:model}). The right panel is the data minus the model; the solid (dashed) contours represent the $+$($-$)$2\sigma$ residuals. The black dotted circles mark the region used to measure the total flux densities (see Figure \ref{fig:alma-cont-cii}).
\label{fig:model_m0}}
\end{figure*}

Here we test whether our current data can rule out a simple rotating disk.
 We model the \cii\ emission as an exponential function that can be described by a thin tilted disk with constant circular velocity and velocity dispersion as described in Section 4 of \cite{venemans2019} and in Appendix C of \cite{neeleman2019}.
We obtain the best fit and uncertainties of the parameters performing a $\chi^2$ minimization using a Markov Chain Monte Carlo approach.
 The best-fit model is shown
 in Figure \ref{fig:model_m0} and consists of a disk with a circular velocity $v_{\rm circ}=57 \pm 14\,\kms$, a velocity dispersion $\sigma_v = 134\pm8\,\kms$, and an inclination angle between the normal to the plane of the galaxy and the line of sight of $i=56^\circ$$^{+4^\circ}_{-6^\circ}$.

In the case of a rotating thin disk geometry, the dynamical mass within a emission region can be expressed as $M_{\rm dyn}/M_\odot=1.16\times 10^5\,v_{\rm circ}^2\,D$ \citep{wang2013}, where $v_{\rm circ}$ is the circular velocity in $\kms$ and $D$ is the disk diameter in kpc. Using the numbers from our best-fit model ($v_{\rm circ}=57\,\kms$ and $D=6.4\,$kpc), it yields a very small dynamical mass of $M_{\rm dyn}=2.4\times 10^9\,M_\odot$, comparable to the mass of the supermassive black hole. This dynamical mass would put \pisco\ about one order of magnitude away from the observed relationships with the black hole mass at lower redshifts and even an outlier between the highest-redshift quasars \citep{kormendy2013,venemans2016}.

\begin{figure*}[ht!]
\plotone{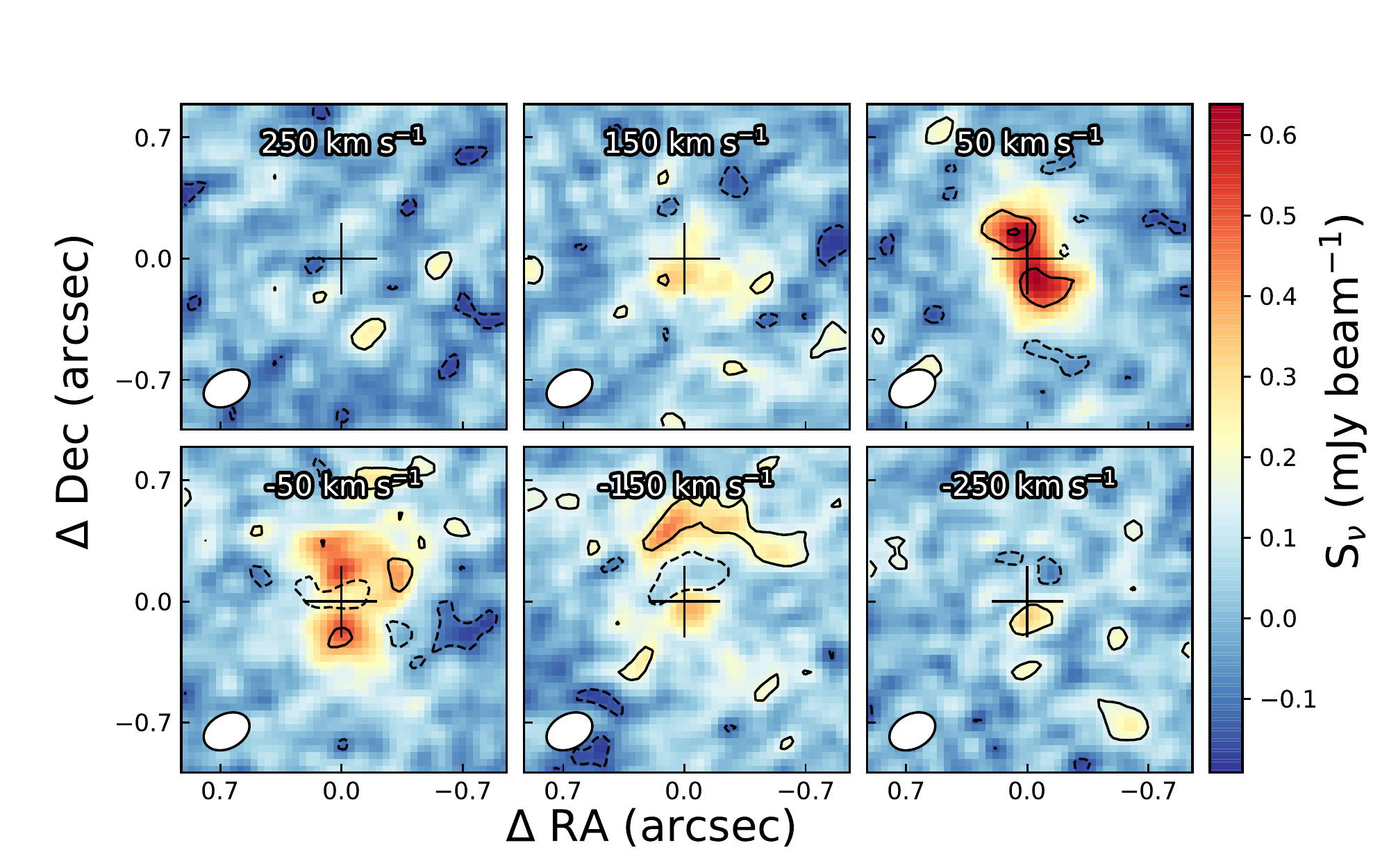}
\caption{
Comparison of \cii\ channel
maps to the best-fit smooth disk model shown in Figure \ref{fig:model_m0}. We show the observations in
color (as in Figure \ref{fig:chmap} but zoomed in by a factor of two) and overplot the  $+$($-$)$2, 4\sigma$ residuals as solid (dashed) contours.
\label{fig:model_chmaps}}
\end{figure*}

At first glance this simple model produces a reasonable fit to our data given the small residuals shown in Figure \ref{fig:model_m0}. However, the fact that $\sigma_v$ is significantly larger than $v_{\rm circ}$ is in contrast to other systems that are well described by a rotating disk (e.g., \citealt{shao2017,neeleman2019}).  If we pay attention to the residuals in the channel maps (Figure \ref{fig:model_chmaps}) there seems to be more substructures that are not well represented by the model.
There are $\pm 2\sigma$ residuals in all channel maps (with some of them spatially coincident), and in one channel map ($50\,\kms$) the model cannot recover the northern structure at $>4\sigma$. These large residuals would not be expected if the simple disk model was an accurate description of the velocity field.

Although a rotating thin disk model is usually assumed to estimate dynamical masses of $z\gtrsim 6$ quasars with much shallower \cii\ data than this work (e.g., \citealt{decarli2018,trakhtenbro2017}), our analysis shows that this assumption might not always be the case. Thus, obtaining deep and high-S/N observations is required to derive robust dynamical host masses.

\subsection{Comparison to simulations}
Our data provide a good test for large cosmological simulations that allow the study and prediction of the observational properties of the first galaxies and supermassive black holes in the universe \citep[e.g.,][]{feng2016,lupi2019}. For example, the brightest quasar in the BlueTides simulation has a similar luminosity and black hole mass at $z=7.54$ to \pisco\ \citep{dimatteo2017, tenneti2019}, but some of the predictions on the host galaxy of the simulated quasar are different from our findings. \cite{ni2018} find that the simulated BlueTides $z=7.54$ quasar produces gas outflows reaching thousands of \kms\ with respect to the systemic redshift, similar to what reported in a quasar at $z=6.4$ \citep{cicone2015}. In contrast, there is no evidence of such an extreme gas outflow in \pisco\ according to our data. In fact, all the \cii\ emission is confined to $\pm 180\,\kms$, with most of the emission within $\pm 100\,\kms$ (see channel maps in Figure \ref{fig:chmap}).
\cite{tenneti2019} find that the host galaxy of the simulated BlueTides $z=7.54$ quasar is fairly featureless with an ellipsoidal shape and a small effective radius of 0.35\,kpc, which contrast to the complex and extended host galaxy of \pisco\ revealed by ALMA (Figures \ref{fig:alma-cont-cii} and \ref{fig:chmap}).

\subsection{Concluding remarks}
Even higher resolution and S/N maps of this quasar are required to better constrain the dynamical mass of this object as well to set meaningful constraints on the clumpiness of its interstellar medium (see discussions in \citealt{hodge2016} and \citealt{gullberg2018}). These observations are within the reach of ALMA and they could resolve the galaxy into more complex and clumpy substructures, perhaps even revealing the formation regions of some of the first star clusters. Furthermore, the fact that the host galaxy of \pisco\ is extended on kiloparsec scales makes it a promising target for exploring its stellar content using the \textit{James Webb Space Telescope}.

\acknowledgments
We thank the referee for constructive feedback and suggestions that improved the quality and presentation of this work.
Ml.N., M.N., B.P.V, and F.W. acknowledge support from ERC Advanced grant 740246 (Cosmic Gas).

This Letter makes use of the following ALMA data: ADS/JAO.ALMA \#2017.1.00396.S. ALMA is a partnership of ESO (representing its member states), NSF (USA) and NINS (Japan), together with NRC (Canada), NSC and ASIAA (Taiwan), and KASI (Republic of Korea), in cooperation with the Republic of Chile. The Joint ALMA Observatory is operated by ESO, AUI/NRAO and NAOJ.

%

\vspace{5mm}
\facility{ALMA.}


\software{
Astropy \citep{astropy2018},
CASA \citep{mcmullin2007},
Matplotlib \citep[][\url{http://www.matplotlib.org}]{hunter2007}
          }

\end{document}